Elastic, electronic, thermodynamic and transport properties of $X$OsSi ($X$=Nb, Ta) superconductors: A first-principles exploration

Enamul Haque and M. Anwar Hossain*

Department of Physics, Mawlana Bhashani Science and Technology University

Santosh, Tangail-1902, Bangladesh

**Abstract**

A first-principles calculation has been performed to study elastic, electronic, thermodynamic, transport and superconducting properties of recently reported osmium based two superconductors, $X$OsSi ($X$=Nb, Ta). We have calculated elastic constants and elastic moduli of $X$OsSi for the first time. The calculated values of bulk, Young's, shear moduli are reasonably larger than the average value obtained from the rule of mixtures of the constituents. NbOsSi and TaOsSi both compounds are found to be relatively hard material, elastically stable and ductile in nature. The obtained directional bulk modulus and shear anisotropic factors indicate that both compounds have high elastic anisotropy. The shear anisotropic factors show higher elastic anisotropy than the percentage anisotropy in these compounds. The Debye temperature and bulk modulus increases with pressure but decreases with temperature as usual for metals. The magnetic susceptibility ($\chi$) of TaOsSi follow the Curie law but NaOsSi do not follow due to its delocalized magnetic moment and electronic specific heat (c) slightly deviates from the linear relationship with temperature. The calculated band structures of $X$OsSi compounds show metallic nature. In both cases d-orbitals have the dominating contribution to the total density of states. The smaller electron-phonon coupling constant implies that $X$OsSi ($X$=Nb, Ta) are weakly coupled superconductors.

**Keywords**: $X$OsSi ($X$=Nb, Ta); Elastic properties; Electronic properties; Thermodynamic properties; Transport properties; Superconductivity.

1.  **Introduction**

Materials with TiNiSi-type crystal structure have gained a huge interest due to their striking magnetic and superconducting properties [1–3]. The equiatomic rare earths, electron deficient transition metals, and silicon form TiNiSi-type silicides [4]. The first osmium based compound in TiNiSi-type silicides is ZrOsSi and the superconducting transition temperature 1.72K is the striking property of this compound [5]. ZrRhSi exhibits highest superconducting transition temperature $T_c$ =10.3K among the TiNiSi-type silicides [6]. Recently, Eckert and co-authors synthesized osmium based two superconductors, NbOsSi and TaOsSi, and found that these compounds exhibit superconductivity below a transition temperature of 3.5K and 5.5K, respectively [7]. The most important structural properties of these two osmium compounds constitute three dimensional [Os-Si] networks and osmium atoms are strongly deformed in the tetrahedral silicon coordination. $X$OsSi ($X$=Nb, Ta) compounds exhibit orthorhombic TiNiSi-type structure. Due to the presence of the refractory metal components Nb and Ta in $X$OsSi, the large cavities are occupied and the interatomic distances are in the range of 2.71-2.74, 2.87-2.96 and 3.26-3.31 Å for $X$-Si, $X$-Os and $X$-$X$ ($X$=Nb, Ta) respectively [7]. The presence of 3d elements in the ternary $TT$Si silicides cause the absence or very low transition temperature. However, TiNiSi type structure rather than the ordered $Fe_2P$ type structures seems to favor the occurrence of superconductivity in the ternary silicides [5].

In this paper, we present a systematic first-principles calculation to study elastic, electronic, thermodynamic, transport, and superconducting properties of recently reported two superconductors: NbOsSi and TaOsSi. We also explore temperature dependent thermodynamic and transport properties. Since some intermetallics exhibits unusual transport properties [8–11], we are interested to investigate the transport properties such as the electrical conductivity, electronic part of the thermal conductivity, Seebeck coefficient, electronic specific heat, and Pauli magnetic susceptibility for $X$OsSi osmium compounds. The Debye temperature is calculated using both the standard Debye model and quasi-harmonic Debye model. We find that the results slightly differ from the value obtained with Debye model. Since there is no experimental value of the Debye temperature, we, therefore, are unable to find the reasonable deviations. The coupling mode of superconductivity is discussed.

## 2. Computational Details

The structural, elastic and electronic properties were investigated by using the full potential linearized augmented plane wave (LAPW) implemented in WIEN2K [12] based on density functional theory (DFT) [13]. A plane wave cut off of kinetic energy $RK_{max}$ =7.0 was selected by convergence tests. A 160 k-points $(9 \times 15 \times 8)$ in the irreducible representations were used for self-consistent calculations. The muffin tin radii were: 2.5, 2.5, 1.81 Bohr for Nb, Os, Si in NbOsSi and 2.5, 2.5, 1.87 Bohr for Ta, Os, Si in TaOsSi, respectively. The elastic properties were calculated by using the IRelast method developed by J. Morteza as implemented in WIEN2K[14]. In all of these calculations, the generalized gradient approximation (GGA) within the Perdew-Burke-Ernzerhof (PBE) scheme was used [15,16]. The pressure and temperature dependent thermodynamic properties were calculated by the modified Gibbs2 program (interfaced with WIEN2K), within the quasi-harmonic model [17]. The semi-classical transport properties were elucidated by solving Boltzmann equation with BoltzTraP program [18,19].The chemical potential equal to the zero temperature Fermi energy was chosen for the transport calculation.

## 3. Result and Discussions

### 3.1. *Structural properties:*

Recently the synthesis and structural characterization of NbOsSi and TaOsSi have reported by C. Benndrof et *al*. [7]. The osmium compounds *X*OsSi (*X*=Nb, Ta) belong to orthorhombic *Pnma* (62) space group and have the internal lattice parameter z (M)=4 [7]. This determines the height of Nb atoms above Os sheets. The structures are similar to TiNiSi in which Os atoms are strongly distorted in the tetrahedral silicon coordination. The crystal structure of *X*OsSi (*X*=Nb,Ta) is shown in the fig.1.

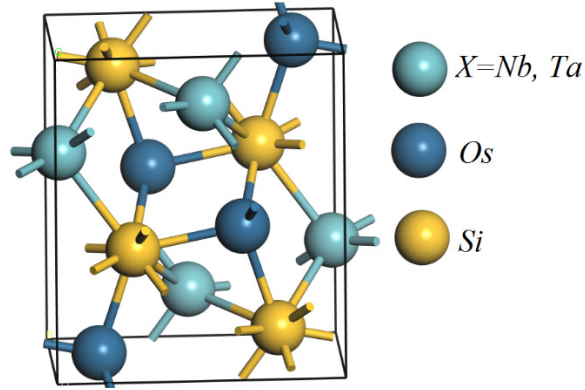

Fig.1: Crystal structure of *X*OsSi (*X*=Nb, Ta).

There are more *X*-Os and Os-Si bondings and in the Os-Si network the "additional" electrons enter into the *X*OsSi structure orbitals cause to be nonbonding. There are four formula units per unit cell and the number of electrons per unit cell are 180 and 236 for NbOsSi and TaOsSi, respectively. The experimental lattice parameters are: a=6.2978, b=3.8872, and c=7.2748 Å for NbOsSi and a=6.268, b=3.8936, and c= 7.2922 Å for TaOsSi [7].

3.2. *Elastic properties*

The mechanical properties give us a deep insight into some fundamental solid-state properties related to the equation of state, anisotropy, hardness, ductility, etc. The elastics properties determine these properties. Elastic properties provide the information about interatomic potentials and phonon spectra. Moreover, they are associated with different thermodynamic parameters such as specific heat, thermal expansion, Debye temperature and Grüneisen parameter. The strength of a crystal depends on the response of the crystal to the external forces. This response is characterized by the bulk modulus, shear modulus, Young's modulus and Poisson's ratio. The elastic constants determine these parameters. There are nine elastic constants for orthorhombic crystal referred as $c_{11}, c_{22}, c_{33}, c_{44}, c_{55}, c_{66}, c_{12}, c_{13}, c_{23}$ and R. Hill described the standard expressions required to calculate these constants [20]. The Reuss shear modulus ($G_R$) and Voigt shear modulus ($G_V$) for orthorhombic crystal can be defined as follows:

$$G_R = \frac{15}{4\left(\frac{1}{c_{11}}+\frac{1}{c_{22}}+\frac{1}{c_{33}}\right)-4\left(\frac{1}{c_{12}}+\frac{1}{c_{13}}+\frac{1}{c_{23}}\right)+3\left(\frac{1}{c_{44}}+\frac{1}{c_{55}}+\frac{1}{c_{66}}\right)} \qquad (1)$$

and

$$G_V = \frac{1}{5}\left[\frac{1}{3}(c_{11}+c_{22}+c_{33}-c_{12}-c_{13}-c_{23})+(c_{44}+c_{55}+c_{66})\right] \qquad (2)$$

The Reuss bulk modulus ($B_R$) and Voigt bulk modulus ($B_V$) for orthorhombic crystal can be expressed as

$$B_R = \frac{1}{\left(\frac{1}{c_{11}}+\frac{1}{c_{22}}+\frac{1}{c_{33}}\right)+2\left(\frac{1}{c_{11}}+\frac{1}{c_{11}}+\frac{1}{c_{11}}\right)} \qquad (3)$$

and

$$B_V = \frac{1}{9}[(c_{11}+c_{22}+c_{33})+2(c_{12}+c_{13}+c_{23})] \qquad (4)$$

Then we can use Hill average scheme to find the bulk ($B$) and shear modulus ($G$) [21]. Other elastic parameters such as Young's modulus, $E = \frac{9BG}{3B+G}$ and Poisson's ratio, $v = \frac{3B-2G}{2(3B+G)}$ are also calculated. The calculated elastic constants are presented in Table 1.

Table 1: Calculated elastic constant for $X$OsSi ($X$=Nb, Ta) in (GPa).

|  | $C_{11}$ | $C_{22}$ | $C_{33}$ | $C_{44}$ | $C_{55}$ | $C_{66}$ | $C_{12}$ | $C_{13}$ | $C_{23}$ |
|---|---|---|---|---|---|---|---|---|---|
| NbOsSi | 437.08 | 408.26 | 509.62 | 120.81 | 131.31 | 105.76 | 195.45 | 149.11 | 188.45 |
| TaOsSi | 461.78 | 431.31 | 534.84 | 128.25 | 141.42 | 113.44 | 204.16 | 156.27 | 196.29 |

The necessary and sufficient conditions for mechanical stability for orthorhombic structure can be written as [22]

$$\begin{cases} c_{11} > 0;\ c_{11}c_{22} > c_{12}^2 \\ c_{11}c_{22}c_{33}+2c_{12}c_{13}c_{23}-c_{11}c_{23}^2-c_{22}c_{13}^2-c_{33}c_{12}^2 > 0 \\ c_{44} > 0;\ c_{55} > 0;\ c_{66} > 0 \end{cases} \qquad (5)$$

From the Table 1 we see that the calculated elastic constants satisfy all conditions (Eq. 5) for elastic stability. Thus, NbOsSi and TaOsSi crystals are elastically stable.

The evaluated polycrystalline elastic moduli and Poisson's ratio of these osmium compounds are presented in the Table 2.

Table 2: Obtained elastic moduli in (GPa) and Poisson's ratio.

|        | $B_R$  | $B_V$  | $B_H$  | $G_R$  | $G_V$  | $G_H$  | $E$    | $\nu$ | $B_H/G_H$ | Ref. |
|--------|--------|--------|--------|--------|--------|--------|--------|-------|-----------|------|
| **NbOsSi** | 268.15 | 268.99 | 268.57 | 123.13 | 126.37 | 124.75 | 324.08 | 0.298 | 2.15 | This |
| **TaOsSi** | 281.57 | 282.38 | 281.97 | 131.44 | 134.70 | 133.07 | 344.96 | 0.296 | 2.12 | This |
| **Nb**  | -      | -      | 143    | -      | -      | 37.5   | 103    | 0.380 | 3.81 | [23] |
| **Ta**  | -      | -      | 206    | -      | -      | 69     | 186    | 0.350 | 2.99 | [23] |
| **Os**  | -      | -      | 415.4  | -      | -      | 271    | 667.8  | 0.232 | 1.53 | [24] |
| **Si**  | -      | -      | 98     | -      | -      | 67     | 163    | 0.220 | 1.46 | [25] |

The bulk modulus (*B*) describes the resistance to fracture. The shear modulus (*G*) describes the resistance to plastic deformation. The ratio of the bulk modulus to shear modulus (*B*/*G*) of polycrystalline phases determines the ductile or brittle nature of a compound [26]. For Pugh ratio (*B*/*G*) ≤ 1.75, a material will be brittle. Otherwise, it will be ductile. It is noted (Table 2) that Pugh ratio for NbOsSi and TaOsSi are 2.15 and 2.12, respectively. Therefore, both crystals are ductile in nature. The Poisson's ratio describes the stability of a compound against shear modulus. The calculated Poisson's ratio of NbOsSi and TaOsSi are given in Table 2. It is noted that Poisson's ratio for NbOsSi and TaOsSi are slightly smaller than the values obtained for other intermetallics [27]. These values indicate that NbOsSi and TaOsSi are relatively stable against shear. The bulk, Young's and shear moduli of NbOsSi and TaOsSi, given in Table-2, are substantially larger than that of the individual constituents Nb, Ta, Si but smaller than that of Os, respectively. However, these elastic moduli of both compounds are larger for the average value obtained from the rule of mixtures and also close to our calculated values. The smaller elastic moduli of NbOsSi and TaOsSi compared to bulk Os are due to the strong deformation of Os in the tetrahedral coordination and weak directional bonding between *X* (*X*=Nb, Ta) and Os. The large elastic moduli of NbOsSi and TaOsSi compared to Nb, Ta, Si, are due to the strong directional bonding between *X* (*X*=Nb, Ta) and Si. The Poisson's ratio of NbOsSi and TaOsSi are smaller than that of the constituents Ta and Nb but larger than that of Os and Si. The Poisson's ratio determines how much volume will change during uniaxial deformation. The larger values of Poisson's ratio means the smaller change in volume and vice versa. For ν=0.5, there will be no change in volume during elastic deformation. Since Poisson's ratios for NbOsSi and TaOsSi are relatively small, large volume change will occur due to elastic deformation. Poisson's ratio can also provide more insight into the nature of bonding forces than other elastic parameters [28]. The lower limit for central forces is ν=0.25 and the upper limit is ν=0.5. The upper limit of central forces means the infinite anisotropy [29]. The calculated Poisson's ratios for NbOsSi and TaOsSi are 0.298 and 0.296, respectively indicate that the interatomic forces in both compounds are central (but weak) and there exists finite elastic anisotropy. The ratios $C_{23}/C_{44}$, $C_{31}/C_{55}$ and $C_{12}/C_{66}$ derived from the Cauchy relations [30] are found to deviates from unity. This deviation from Cauchy relations supports the above view point.

Since both crystals are ordered Fe$_2$P (TiNiSi) type and thus they are intermetallic compounds [31]. Both the thermal expansion coefficient anisotropy and elastic anisotropy are responsible for nucleating microcracks in the ceramics [32]. Therefore, it is important to study elastics anisotropy to reveal the nature of microcracks and with the hope to find the possible mechanism to increase the durability of these compounds. Thus, the proper explanation of elastic anisotropy is important in material and related engineering fields or in crystallography. In NbOsSi and TaOsSi atoms are bonded in different planes. The shear anisotropic factors measure the degree of anisotropy in these planes. Between $<011>$ and $<010>$ directions the shear planes is {100}. For this plane, the shear anisotropic factor can be written as [20]

$$A_1 = \frac{4C_{44}}{C_{11}+C_{33}-2C_{13}} \tag{6}$$

For the shear planes {010} (between the $<101>$ and $<001>$ directions) the shear anisotropic factor can be written as

$$A_2 = \frac{4C_{55}}{C_{22}+C_{33}-2C_{23}} \tag{7}$$

and for the shear planes {001} (between the $<110>$ and $<001>$ directions) it can be written as

$$A_3 = \frac{4C_{66}}{C_{11}+C_{22}-2C_{12}} \tag{8}$$

The elastic anisotropy of an orthorhombic crystal comes from linear bulk modulus as well as shear anisotropy. The compressibility anisotropic factors can be written as follows:

$$A_{B_a} = \frac{(C_{11}-C_{12})(C_{33}-C_{13})-(C_{23}-C_{13})(C_{11}-C_{13})}{(C_{33}-C_{13})(C_{22}-C_{12})-(C_{13}-C_{23})(C_{12}-C_{23})} = \epsilon \qquad (9)$$

and

$$A_{B_c} = \frac{\epsilon}{\frac{(C_{22}-C_{12})(C_{11}-C_{13})-(C_{11}-C_{12})(C_{23}-C_{12})}{(C_{22}-C_{12})(C_{33}-C_{13})-(C_{12}-C_{23})(C_{13}-C_{23})}} = \alpha \qquad (10)$$

Here the anisotropies of bulk modulus $A_{B_a}$ and $A_{B_c}$ is along **a**-axis and **c**-axis with respect to **b**-axis. The upper bound to the bulk modulus ($B_{unrelax}$) can easily be calculated by using the following expression,

$$B_{unrelax} = \frac{C_{11}+2C_{12}\epsilon+C_{22}\epsilon^2+2C_{13}\alpha+C_{33}\alpha^2+2C_{23}\epsilon\alpha}{9} = \frac{\beta}{9} \qquad (11)$$

Using the above relation, we can also calculate the directional bulk modulus, i.e., the bulk modulus along three crystallographic axes **a**, **b**, and **c**.

$$B_a = \frac{\beta}{1+\epsilon+\alpha}, \quad B_b = \frac{B_a}{\epsilon} \quad \text{and} \quad B_c = \frac{B_a}{\alpha} \qquad (12)$$

The calculated directional bulk modulus are given in Table 4. However, Chung and Buessem introduced the concept of percentage anisotropy. This measures the elastic anisotropy and defined as [33]

$$A_G = \frac{G_V-G_R}{G_V+G_R} \quad \text{and} \quad A_B = \frac{B_V-B_R}{B_V+B_R} \qquad (13)$$

Table 3: Calculated unrelaxed bulk modulus, bulk modulus along the orthorhombic crystallographic axes, **a**, **b**, **c** ( in GPa ).

|  | $B_{unrelux}$ | $B_a$ | $B_b$ | $B_c$ |
| --- | --- | --- | --- | --- |
| NbOsSi | 234.3 | 751.9 | 764.1 | 917.0 |
| TaOsSi | 248.8 | 794.0 | 802.0 | 956.7 |

Another important parameter is the hardness and can be calculated by using the following equation [34]

$$H_V = 2\left(\left(\frac{G}{B}\right)^2 G\right)^{0.585} - 3 \tag{14}$$

The obtained anisotropic factors and Vicker's hardness for *X*OsSi are provided in the Table 4.

Table 4: Calculated elastic anisotropic factors, $A_B$, $A_G$ (in %), and Vicker's hardness (GPa).

|        | $A_1$ | $A_2$ | $A_3$ | $A_B$ | $A_G$ | $A_{B_a}$ | $A_{B_c}$ | $H_v$ |
|--------|-------|-------|-------|-------|-------|-----------|-----------|-------|
| NbOsSi | 0.75  | 0.97  | 0.93  | 0.15  | 1.3   | 0.98      | 1.20      | 10.73 |
| TaOsSi | 0.75  | 0.99  | 0.94  | 0.14  | 1.2   | 0.99      | 1.19      | 11.52 |

For $A_1$, $A_2$, $A_3$ equal to 1, a crystal is referred as an isotropic crystal. A smaller or greater value from 1 describes the degree of elastic anisotropy. For percentage anisotropy, the zero value indicates a crystal to isotropic and the unit value indicates the largest possible anisotropy in the crystal. The directional bulk moduli (in Table 3) are larger and shear anisotropic factors (in Table 4) are smaller for TaOsSi than NbOsSi, respectively. These values show that both compounds have high elastic anisotropy. It is interesting to note that shear anisotropic factors show higher anisotropy than percentage anisotropy.

The nature of deformation and atomic bonding of transition metal silicides are related to the elastic constants $C_{11}$ and $C_{33}$. It is noted from Table-1 that $C_{33} > C_{11}$ for both compounds. Therefore, The atomic bonds between nearest neighbors along {100} planes are weaker than these along {001} planes. Tanaka *et al.* introduced a concept that bulk modulus describes the average bond strength and shear modulus describes the resistance to an external forces to change the bond angle [35]. Thus, the ratio of shear modulus to bulk modulus (G/B) measures the relative directionality of atomic bonds in a crystal. For NbOsSi and TaOsSi, the values of G/B are larger than Nb and Ta but smaller than Si and Os indicating that the directionality in the bonding of both compounds is stronger than Nb and Ta but weaker than Os and Si, respectively. It is noted from Table 3 that the directional bulk moduli along three crystallographic axes are larger than unreluxed bulk modulus. However, the unrelaxed bulk modulus is also smaller than the Voigt bulk modulus. The value of Vickers's hardness $H_V$ indicates that both compounds are relatively hard material and TaOsSi is harder than NbOsSi. The elastic constants can also be used to calculate Debye temperature. Debye temperature of solid is a measure of the cut off frequency and is proportional to the Debye sound velocity and can be expressed as:

$$\theta_D = \frac{h}{2\pi \kappa_B} \left(\frac{6\pi^2 N}{V}\right)^{1/3} v_D \qquad (15)$$

where $\kappa_B$, $V$, $N$ and $v_D$ are the Planck constant, volume of solid, number of atoms and Debye sound velocity. For real solids the Debye velocity of elastic wave is related with longitudinal wave velocity ($v_l$) and transverse wave velocity ($v_t$) as $\frac{1}{v_D^3} = \frac{1}{3}\left(\frac{1}{v_l^3} + \frac{1}{v_t^3}\right)$. The calculated elastic wave velocities and Debye temperature for $X$OsSi ($X$=Nb, Ta) are presented in Table 5.

Table 5: Obtained thermodynamic parameters from elastic properties at 0 GPa: Transverse elastic wave ($v_t$), longitudinal elastic wave ($v_l$) and Debye wave velocity ($v_D$) in m/s, Debye Temperature ($\theta_D$) in K for $X$OsSi ($X$=Nb, Ta).

|         | $v_l$ | $v_t$ | $v_D$ | $\theta_D$ |
|---------|-------|-------|-------|------------|
| NbOsSi  | 3278  | 6121  | 3661  | 444        |
| TaOsSi  | 2988  | 5552  | 3336  | 404        |

The Debye temperature is directly proportional to the square root of melting point [36] and hence it is expected that the melting point of NbOsSi will be higher than TaOsSi.

### 3.3. Electronic properties

The GGA energy band structures for $X$OsSi ($X$=Nb, Ta) are illustrated in Fig. 2 (a) and (b) (only the d-characters of Nb, Os, Ta and Os are plotted since other characters are very small near the Fermi level) for NbOsSi and TaOsSi, respectively. The basic characteristics of band structures of NbOsSi and TaOsSi are similar. However, the d-bands are narrower for Nb compounds than Ta. The majority and minority bands are overlapped around the Fermi level and metallic character is observed. There are ten d-states (five for Nb or Ta and five for Os) near the Fermi level which have the most contribution to the density of states. There is a small pseudogap below the Fermi level. All these bands are non-dispersive and only 66, 67 and 56, 57-th bands for NbOsSi cross the Fermi level.

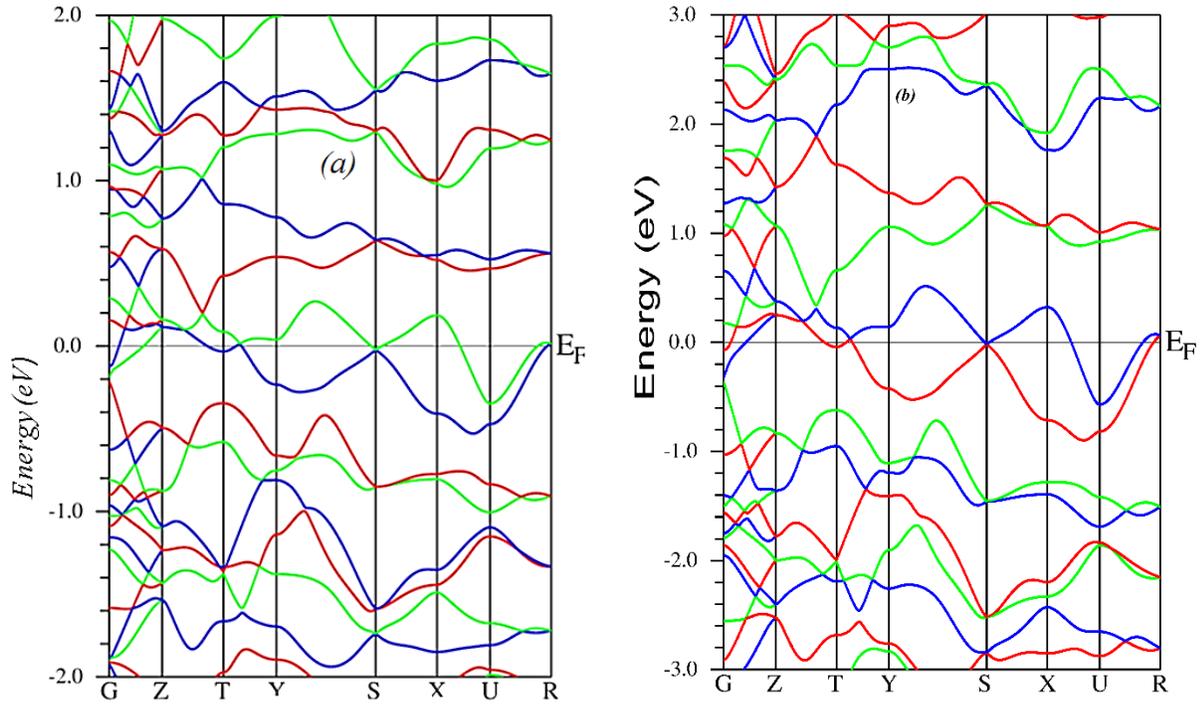

Fig. 2: Band structure of (a) NbOsSi and (b) TaOsSi.

The calculated electronic density of states of NbOsSi and TaOsSi are illustrated in the Fig.3 (a) and (b). The calculated total density of states for NbOsSi and TaOsSi are 7.30 and 6.97 electrons/eV, respectively. It is also observed that Nb-4d, Os-5d orbitals for NbOsSi and Ta-5d, Os-5d orbitals for TaOsSi have dominating contribution to the total density of states. In both cases Si-3p has the minimum contribution to the total density of states at the Fermi level.

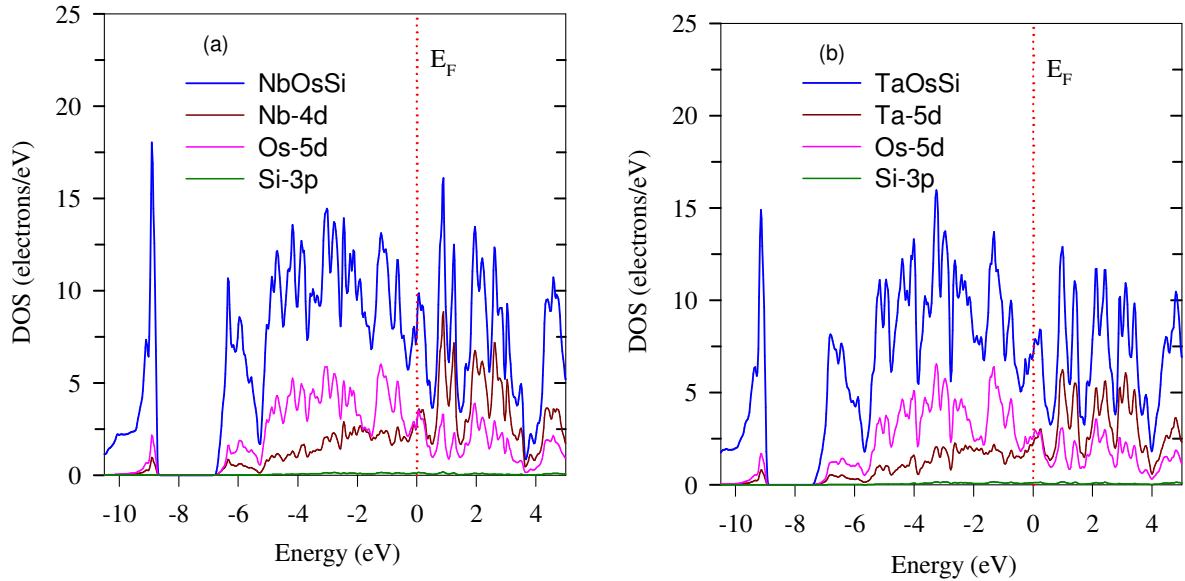

Fig.3: Density of states (DOS) of : (a) NbOsSi and (b) TaOsSi. (The dotted vertical line is the Fermi level set at zero energy).

3.4. *Thermodynamic properties*

Thermodynamic properties are related to the stability, durability of a crystal. Thermal properties such as specific heat and Debye temperature are directly related with the superconducting transition temperature. The TiNiSi-type structures exhibit some interesting thermodynamic properties such as quadratic temperature dependence of resistivity, enhanced electronic specific heat coefficient [37]. The Debye temperature decreases with increasing temperature for intermetallic compounds [38]. Thus it is interesting to study thermal properties of NbOsSi and TaOsSi. The equilibrium energy, bulk modulus, volume, and other required parameters to calculate thermodynamic quantities were obtained by volume optimization. The temperature range was automatically set from 0 to 386.78 K. The Murnaghan equation of state was used to fit the energy versus volume and thus using standard thermodynamic relations the macroscopic thermodynamic quantities as a function of pressure (P) and temperature (T) were calculated. The temperature and pressure dependent thermodynamic parameters are illustrated in Fig. 4.

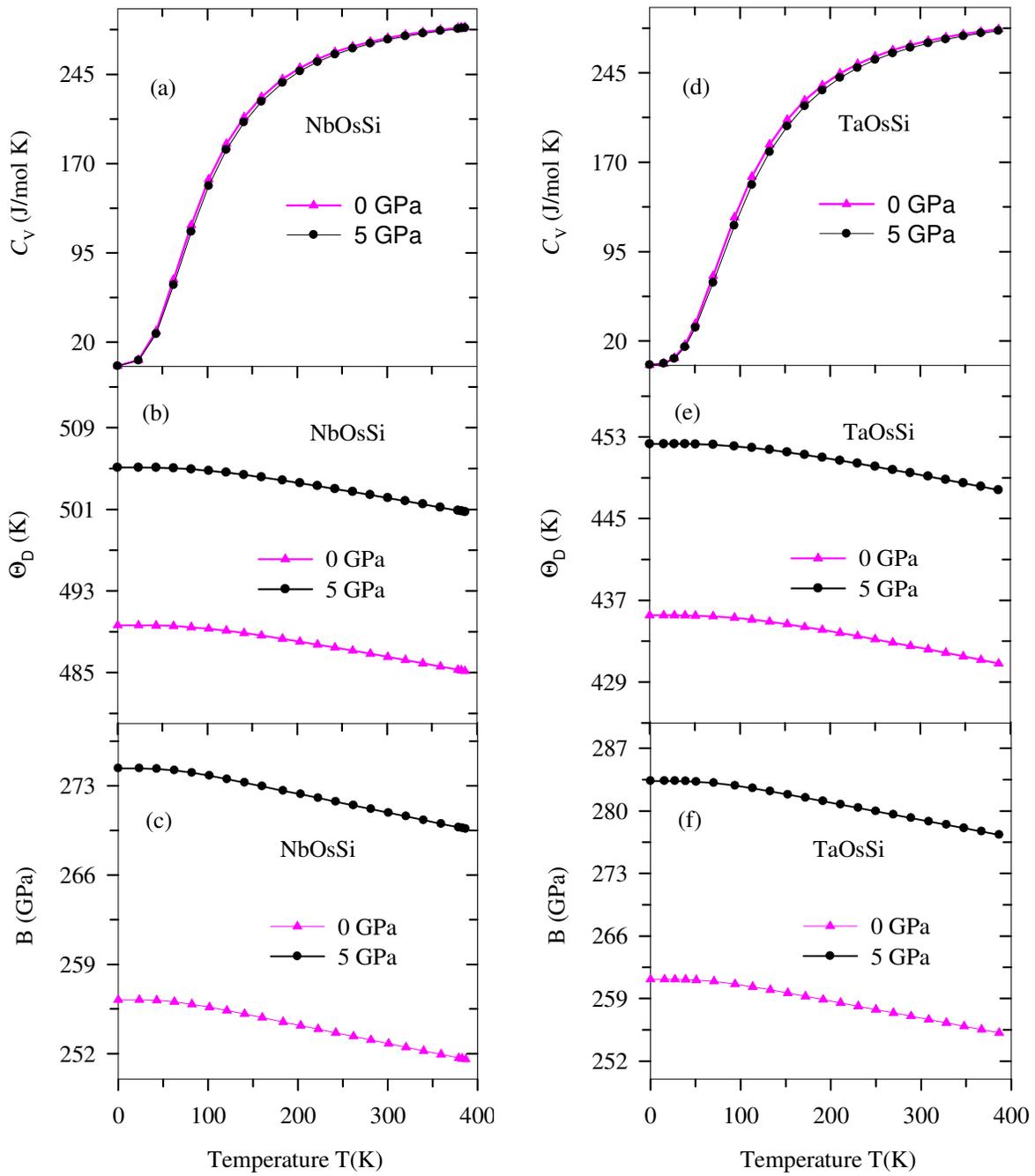

Fig. 4: Temperature and pressure dependence of bulk modulus, Debye temperature and specific heat at constant volume for NbOsSi and TaOsSi.

Fig. 4 (a) and (d) illustrates the variations of heat capacity ($C_V$) with temperature at 0 and 5 GPa for $X$OsSi ($X$=Nb, Ta). It is clear that there is no significant pressure effect on $C_V$ within the temperature range studied. As temperature increases, $C_V$ approaches the Dulong-Petit limit, as expected for any solids [39]. Similar results are shown in Fig. 4(d) for TaOsSi. It is clear that at 0 GPa and 300K, the values of $C_V$ for $X$OsSi are approximately 275.6 J/mol K. The variations of Debye temperature with temperature for $X$OsSi are illustrated in the Fig. 4 (b), (e). It is found for both compounds that Debye temperature below 39K are almost constant and above 39K decreases almost linearly. The Debye temperature increases rapidly with the increase in pressure as it is expected. Since Debye temperature is inversely related to the vibrational frequency, these results show that vibrational frequency increases with the increasing temperature but decreases due to the increase of pressure. Thus phonon contribution increases with increasing pressure but decreases with increasing temperature. The degree of resistance of a material is determined by the bulk modulus. A material will be more resistive when the bulk modulus is larger. Fig. 4 (c) and (f) show the variations of the bulk modulus with temperature at 0 and 5GPa for NbOsSi and TaOsSi, respectively and we see that the bulk modulus increases with increasing pressure but decreases with increasing temperature. Thus the degree of resistance of $X$OsSi decreases with increasing temperature but increases with increasing pressure and this behavior is responsible for the change in volume due to pressure and temperature effect. However, TaOsSi is more resistive material than NbOsSi, as it is seen from Fig. 4 and Table 2. We found that bulk moduli for $X$OsSi compounds at 0 GPa obtained from Gibbs2 program are very close to that values obtained from elastic properties using standard Debye model.

### 3.5. *Transport properties*

The intermetallic TiNiSi type structures may exhibit magnetic and unconventional transport properties [40–43]. It was found that intermetallic CeRhSb (TiNiSi-type) show some interesting and unusual transport properties [44]. It is, therefore, interesting to study the transport properties of NbOsSi and TaOsSi. The calculated electronic specific heat (c), conductivity ($\sigma/\tau$), electronic thermal conductivity ($\kappa_e/\tau$), Seebeck coefficient ($S$) at different temperature are plotted in the Fig. 5.

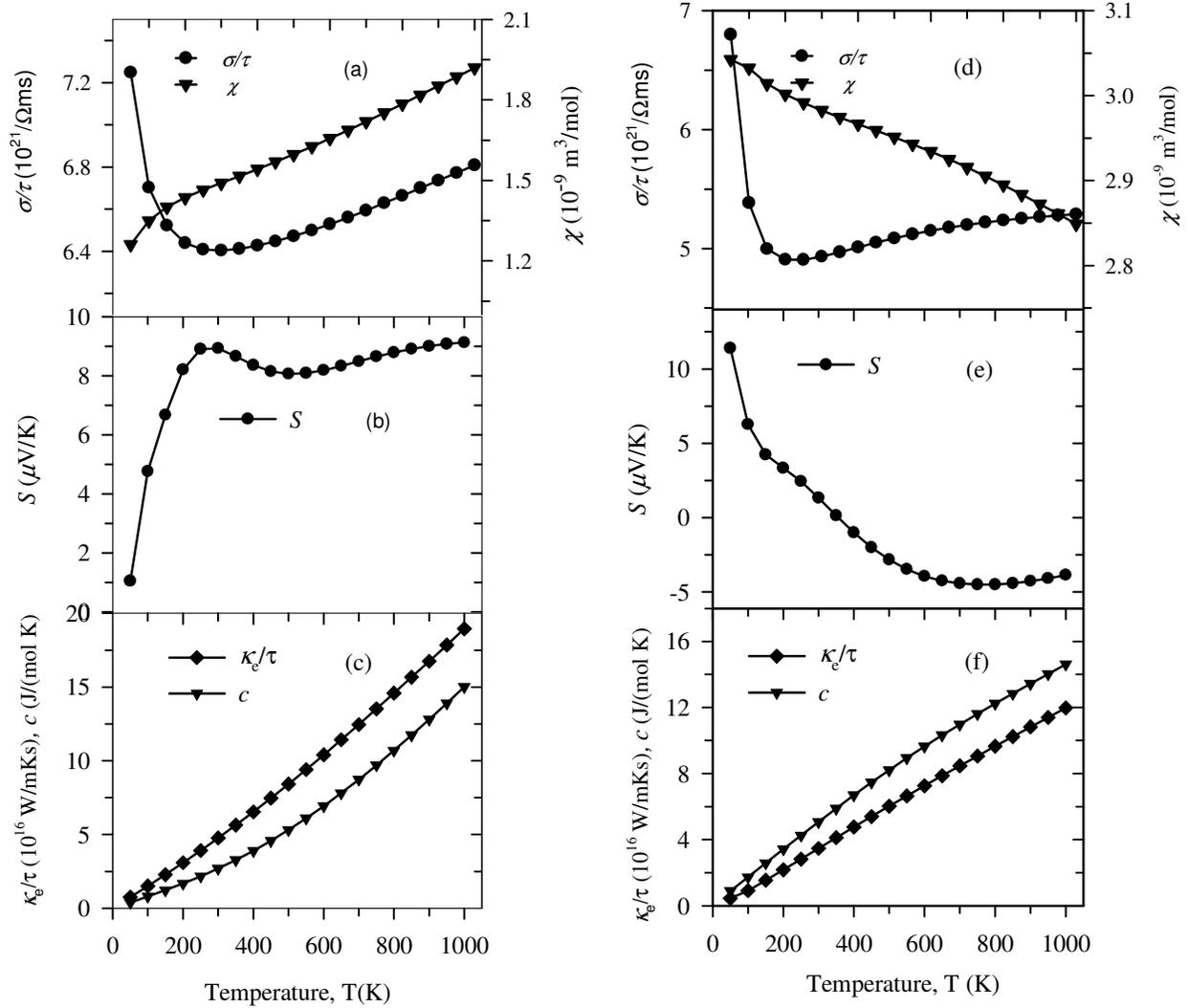

Fig. 5: Temperature dependent transport properties: (a)-(c) for NbOsSi and (d)-(f) for TaOsSi.

It is clear that the electrical conductivity ($\sigma/\tau$) decreases rapidly for temperature T ≤ 200 K as shown in Fig. 5 (a) and (d). This sharp decrease $\sigma/\tau$ indicates the metallic nature. As temperature increases (from 200k) $\sigma/\tau$ increases slowly. This slow increase of $\sigma/\tau$ signifies semiconducting nature of both compounds. The Seebeck coefficient (*S*) increases linearly with temperature upto 250K and after then it remains almost constant for NbOsSi as illustrated in Fig. 5 (b). This indicates that the NbOsSi is p-type material (as *S* is positive). However, *S* decreases rapidly with increasing temperature and remains positive upto 250K for TaOsSi as shown in Fig. 5 (e). Thus TaOsSi crystal remains p-type materials upto 250K and after this temperature, it becomes n-type materials. The Pauli magnetic susceptibility ($\chi$) unusually increases almost in a linear fashion with temperature for NbOsSi as shown in Fig. 5 (a). The Curie law [45] is not valid, i.e., $\chi \propto 1/T$ is not hold for NbOsSi. Thus, the magnetic moment of NbOsSi is delocalized at the atomic sites and within this region and there is no interaction between neighboring magnetic moments [46]. However, the situation is completely different for TaOsSi as shown in Fig. 5 (d). For this compound, $\chi$ decreases linearly with temperature, thus Curie law holds. Therefore, the magnetic moment of TaOsSi is localized at the atomic sites and interacts with neighboring magnetic moments. The electronic heat capacity (c) increases linearly with temperature for TaOsSi but for NbOsSi it increases slightly non-linearly as illustrated in Fig. 5 (c) and (f). Thus, for TaOsSi, the linear relation between the electronic heat capacity and temperature holds, i.e., $c = \gamma T$, where $\gamma$ is Sommerfeld constant [47,48]. For NbOsSi, this relationship slightly deviates. When electrons are excited to upper empty state, there is a smearing of Fermi level. The electronic heat capacity comes from the combination of this two contributions. These behaviors might cause smaller superconducting transition temperature for NbOsSi (3.5K) compared to TaOsSi (5.5K). For both compounds the electronic thermal conductivity increase with temperature and this implies their metallic nature as shown in Fig. 5(c) and (f).

3.6. *Superconducting properties*

Superconductivity is a fascinating phenomenon and it may arise due to electron-phonon interaction, strong spin-orbit interaction, electron-electron pairing. We only consider here the electron-phonon interaction. The electron-phonon coupling constant (EPC) can easily be calculated if we know experimental specific heat coefficient (γ). Since the Sommerfeld coefficient for non-interacting particles is proportional to the DOS at the Fermi level, thus the electronic specific heat coefficient (γ) is

$$\gamma_c = \frac{1}{3}\pi^2 \kappa_B^2\, N(E_F) \tag{15}$$

Then the EPC can be obtained as [49]

$$\lambda = \frac{\gamma_{exp}}{\gamma_c} - 1 \tag{16}$$

However, there is no experimental value of electronic specific heat (γ) for *X*OsSi. Thus we cannot use equations (15) and (16). Alternatively, the EPC can be calculated from the inverted McMillan formula [50]

$$\lambda_{EPC} = \frac{1.04 + \mu^* \ln\left(\frac{\theta_D}{1.45 T_C}\right)}{(1 - 0.62\mu^*)\ln\left(\frac{\theta_D}{1.45 T_C}\right) - 1.04} \tag{17}$$

The Coulomb pseudopotential can be obtained by using Bennemann-Garland formula [51,52]

$$\mu^* = 0.13 \frac{N(E_F)}{1 + N(E_F)} \tag{18}$$

The reason for writing the prefactor 0.13 is that the value of μ* should be consistent with the experimental volume. Here the DOS at the Fermi level is expressed in per unit cell. The calculated superconducting parameters are presented in Table 6.

Table 6: Calculated superconducting parameters.

|  | $N(E_F)$ | $\mu^*$ | $T_c$ (K) | $\theta_D$ (K) | $\lambda_{EPC}$ |
|---|---|---|---|---|---|
| NbOsSi | 7.30 | 0.1143 | 3.5[a] | 444 | 0.50 |
| TaOsSi | 6.97 | 0.1136 | 5.5[a] | 404 | 0.57 |

We obtained the electron-phonon coupling constants, $\lambda_{EPC}$ for NbOsSi and TaOsSi are 0.50 and 0.57, respectively. These small values of $\lambda_{EPC}$ indicate that both compounds are weakly coupled, i.e., there is a weak interaction between electrons and phonons. This contribution mainly comes from the d-orbitals of *X* and Os for *X*OsSi (*X*=Nb, Ta).

## 4. Conclusion

In this paper we have performed a comprehensive first-principles calculation to study structural, elastic, electronic, thermodynamic, transport and superconducting properties of two osmium-based superconductors, NbOsSi and TaOsSi within the generalized gradient approximation using a full potential linearized augmented plane wave (LAPW) method. We calculated elastic constants and elastic moduli using the IRelast method implemented in WIEN2k. The average value obtained from the rule of mixture using the measured values of bulk, Young's and shear moduli for individual constituents reasonably smaller than our calculated values of bulk, Young's, and shear modulus of NbOsSi and TaOsSi but very close to our calculated value. The superconducting NbOsSi and TaOsSi crystals are found to be relatively hard material, elastically stable and ductile in nature. The shear anisotropic factors $A_1$, $A_2$, $A_3$, show higher elastic anisotropy than the percentage anisotropy $A_G$, $A_B$ in $X$OsSi compounds. We calculated Debye temperatures from elastic constants. The temperature and pressure dependent thermodynamic properties are obtained by solving the equation of state (EOS). The Debye temperature and bulk modulus increases with pressure but decreases with temperature as usually show by metals. The pressure has a small effect on the heat capacity ($C_V$) and after certain temperature limit pressure effect totally vanished. The magnetic susceptibility ($\chi$) of NbOsSi does not follow the Curie law due to its delocalized magnetic moment and electronic specific heat (c) slightly deviates from the linear relationship with temperature. But for TaOsSi, these relationship is well obeyed. These behaviors may be responsible for low superconducting transition temperature of NbOsSi (3.5K) than TaOsSi (5.5K). The calculated band structures show that all bands are non-dispersive and metallic in nature near the Fermi level. The calculated density of states reveal that the d-orbitals have the highest contribution to the total density of states for both compounds. The electron-phonon coupling constants for NbOsSi and TaOsSi are 0.50 and 0.57, respectively and the smaller values indicate they are weakly coupled superconductors. The detailed mechanism of the origin of superconductivity in $X$OsSi beyond the scope of this paper is in progress.